\documentclass[prd,showpacs,preprintnumbers,amsmath,amssymb,nofootinbib]{revtex4}
\usepackage{cancel}
\usepackage{booktabs}
\usepackage{multirow}
\usepackage{slashed}
\usepackage{graphicx}
\usepackage{rotating}
\usepackage[colorlinks,
            citecolor=blue,
            anchorcolor=red,
            menucolor=red,
            linkcolor=red,
            filecolor=red,
            runcolor=red,
            urlcolor=blue,
            frenchlinks=red]{hyperref}

\begin{document}
\title{$P_c(4457)^+$, $P_c(4440)^+$, and $P_c(4312)^+$: molecules or compact pentaquarks?}
\author{Jian-Bo Cheng$^{1}$}
\author{Yan-Rui Liu$^{1}$}
\email{yrliu@sdu.edu.cn} \affiliation{ $^1$School of Physics, Shandong University, Jinan 250100, China}

\date{\today}
\begin{abstract}
In a chromomagnetic model, we analyse the properties of the newly observed $P_c(4457)^+$, $P_c(4440)^+$, and $P_c(4312)^+$ states. We estimate the masses of the $(uud)_{8_c}(c\bar{c})_{8_c}$ and $(uds)_{8_c}(c\bar{c})_{8_c}$ pentaquark states by considering the isospin breaking effects. Their values are determined by calculating mass distances from the $\Sigma_c^{++}D^-$ and $\Xi_c^{\prime+}D^-$ thresholds, respectively. It is found that the isospin breaking effects on the spectrum are small. From the uncertainty consideration and the rearrangement decay properties in a simple model, we find that it is possible to assign the $P_c(4457)^+$, $P_c(4440)^+$, and $P_c(4312)^+$ as $J^P=3/2^-$, $1/2^-$, and $3/2^-$ pentaquark states, respectively. The assignment in the molecule picture can be different, in particular for the $P_c(4312)^+$. The information from open-charm channels, e.g. ${\cal B}[P_c\to\Sigma_c^{++}D^-]/{\cal B}[P_c\to J/\psi p]$, will play an important role in distinguishing the inner structures of the $P_c$ states. Discussions and predictions based on the calculations are also given.
\end{abstract}


\maketitle
\section{Introduction}\label{sec1}

The LHCb Collaboration observed two pentaquark-like $P_c(4380)$ and $P_c(4450)$ states in the invariant mass distribution of $J/\psi p$ in the $\Lambda_b^0$ decay in 2015 \cite{Aaij:2015tga,Aaij:2016phn,Aaij:2016ymb}. In the literature, there are pioneering works about hidden-charm type pentaquarks before the observation  \cite{Wu:2010jy,Wu:2010vk,Wang:2011rga,Yang:2011wz,Wu:2012md} and lots of heated discussions on their structures motivated by this observation. These two $P_c$ states have been widely discussed in the $\Sigma_c\bar{D}^*$, $\Sigma_c^*\bar{D}$, or $\Sigma_c^*\bar{D}^*$ molecule picture, diquark-diquark-antiquark picture, diquark-triquark picture, compact pentaquark picture, and nonresonance picture. One may consult Refs. \cite{Chen:2016qju,Liu:2019zoy,Richard:2016eis,Lebed:2016hpi,Esposito:2016noz,Guo:2017jvc,Ali:2017jda,Olsen:2017bmm,Karliner:2017qhf} for details about relevant investigations existing in recent years. Very recently, the LHCb announced the new observation of three narrow pentaquark-like states in the $J/\psi p$ channel with their updated data for $\Lambda_b^0 \to J/\psi p K^-$ \cite{Aaij:2019vzc}. The measured masses and widths are
\begin{eqnarray}
M(P_c(4312)^+)&=&4311.9\pm0.7^{+6.8}_{-0.6}\text{ MeV}, \qquad\Gamma(P_c(4312)^+)=9.8\pm2.7^{+3.7}_{-4.5}\text{ MeV},\nonumber\\
M(P_c(4440)^+)&=&4440.3\pm1.3^{+4.1}_{-4.7}\text{ MeV}, \qquad\Gamma(P_c(4440)^+)=20.6\pm4.9^{+8.7}_{-10.1}\text{ MeV},\nonumber\\
M(P_c(4457)^+)&=&4457.3\pm0.6^{+4.1}_{-1.7}\text{ MeV}, \qquad\Gamma(P_c(4457)^+)=6.4\pm2.0^{+5.7}_{-1.9}\text{ MeV}.
\end{eqnarray}
The previously observed $P_c(4450)$ actually consists of the latter two states.

These three newly observed $P_c$ states are certainly helpful to deepen our understandings about the nature of strong interactions. Because these states are just below the $\Sigma_c\bar{D}$ or $\Sigma_c\bar{D}^*$ threshold with narrow decay widths, a typical feature expected for molecules, their structures in the molecule picture have been discussed through various approaches, the QCD sum rule method \cite{Chen:2019bip,Zhang:2019xtu}, the one-boson-exchange model \cite{Chen:2019asm}, the chiral effective field theory (EFT) \cite{Meng:2019ilv}, a contact-range EFT \cite{Liu:2019tjn}, a quasipotential Bethe-Salpeter equation approach \cite{He:2019ify}, the quark-level or hadron-level scattering methods \cite{Huang:2019jlf,Guo:2019kdc,Xiao:2019aya}, solving the quark-level 5-body problem \cite{Mutuk:2019snd,Zhu:2019iwm}, studying the decay properties \cite{Guo:2019fdo,Xiao:2019mst} or the multiplet structures \cite{Shimizu:2019ptd}, and analyzing the underlying reaction amplitude \cite{Fernandez-Ramirez:2019koa}. Most investigations favor the molecule picture. A different interpretation is to assign them as hadrocharmonium states \cite{Eides:2019tgv}. Constraints on the branching ratios of $P_c\to J/\psi p$ are discussed in Ref. \cite{Cao:2019kst} while the photoproductions of the $P_c$ states in $\gamma p\to J/\psi p$ are studied in Ref. \cite{Wang:2019krd}.

In fact, the present experimental data did not exclude the possibility that the LHCb $P_c$ states are compact pentaquarks, since information about quantum numbers and decay properties is still lacking. The compact diquark-diquark-antidiquark picture \cite{Ali:2019npk,Wang:2019got} and tightly bound pentaqurk picture \cite{Weng:2019ynv} can also give masses close to the measured values. In this paper, we would like to investigate another possibility, namely, they are compact pentaquark states composed of colored $c\bar{c}$.

Previously, we have systematically estimated the spectra of the $(qqq)_{8_c}(c\bar{c})_{8_c}$ pentaquarks in Ref. \cite{Wu:2017weo} with a chromomagnetic interaction (CMI) model. The masses of $P_c(4380)$ and $P_c(4450)$ fall in the mass region of such states. The mass of the lowest $uudc\bar{c}$ state was found to be consistent with the prediction in the molecule picture given in Ref. \cite{Wu:2010jy}. If one checks for the lowest $udsc\bar{c}$ state, these two pictures also give comparable masses. Note that the obtained masses in the compact pentaquark picture mainly rely on the mass splittings of conventional hadrons. The comparable results indicate that these two pictures probably have some relations in the $qqqQ\bar{Q}$ systems because both calculations exclude the $(c\bar{c})_{1_c}(uud)_{1_c}$ configuration\footnote{In a similar study for the $(QQQ)_{8_c}(q\bar{q})_{8_c}$ systems \cite{Li:2018vhp}, the compact pentaquark picture gives higher masses than the molecule picture. In that case, an observed state around the $\Xi_{cc}D$ threshold can be claimed as a molecule undoubtedly.}. If the $P_c(4312)^+$ is a $\Sigma_c\bar{D}$ molecule state, the short-range $\rho$-meson exchange interaction would be important \cite{Aaij:2019vzc,Karliner:2015ina}. Note that the important short-range interaction also implies that the gluon-exchange interaction should not be negligible\footnote{The proposal of the extended chiral quark model in Ref. \cite{Dai:2003dz} is based on the spirit that the vector-meson exchange interactions can replace part of one-gluon-exchange interactions \cite{Glozman:1995fu,Glozman:1999vd}.}. It is possible to explain the $P_c$ masses in both pictures. However, the decay properties of pentaquarks should be different and the relevant studies are helpful to distinguish their structures. Here, we improve the spectra of $uudc\bar{c}$ and $udsc\bar{c}$ states and explore their rearrangement decay properties by considering the isospin breaking effects. Comparison of decays in this pentaquark picture and in the molecule picture will also be discussed. In understanding the nature of $P_c(4380)$ and $P_c(4450)$ and in predicting other pentaquark states, investigations in this compact picture were also performed in Refs. \cite{Takeuchi:2016ejt,Irie:2017qai,Richard:2017una,Li:2017ghe,Buccella:2018jnt}.

This paper is organized as follows. In Sec. \ref{sec2}, we present the formalism for the study. Then we give our numerical results and discussions in Sec. \ref{sec3} for the $uudc\bar{c}$ pentaquark states. Sec. \ref{sec4} is about the $udsc\bar{c}$ states. We give some discussions and a short summary in the final Sec. \ref{sec5}.

\section{Formalism}\label{sec2}

\subsection{Spectrum}

When calculating the mass splittings between the pentaquarks, we use the color-magnetic interaction (CMI) model \cite{Liu:2019zoy}. The coupling term reads
\begin{eqnarray}\label{Hamiltonian}
H_{CM}&=&-\sum_{i<j}C_{ij}\lambda_i\cdot\lambda_j\sigma_i\cdot\sigma_j,
\end{eqnarray}
where the $i$-th Gell-Mann matrix $\lambda_i$ should be replaced with $-\lambda_i^*$ for an antiquark. The effective coupling parameter $C_{ij}$ between the $i$-th quark component and the $j$-th quark component can be determined from the mass splittings of the known ground hadrons. Their values we will use are shown in Table \ref{parameter}. Here, the isospin breaking effects have been considered.

\begin{table}[!h]
\caption{The effective coupling parameters (units: MeV) determined from the mass differences between ground hadrons. The isospin breaking effects have been considered.}\label{parameter} \centering
\begin{tabular}{cccc}\hline
$C_{uu}=20.01$&$C_{ud}=19.16$&$C_{dd}=18.87$&$C_{uc}=4.03$\\
$C_{dc}=4.05$&$C_{u\bar{c}}=6.66$&$C_{d\bar{c}}=6.59$&$C_{c\bar{c}}=5.30$\\
$C_{us}=12.09$&$C_{ds}=11.86$&$C_{sc}=4.44$&$C_{s\bar{c}}=6.74$\\
\hline
\end{tabular}
\end{table}

The required matrix elements of $H_{CM}$ can be calculated by constructing the flavor-color-spin wave
functions of the ground pentaquark states. We have obtained the expressions of $\langle H_{CM}\rangle$ in Ref. \cite{Wu:2017weo} and do not repeat all the results here.

When one considers the $q_1q_2q_3q_4\bar{q}_5=uudc\bar{c}$ pentaquark states, the CMI matrices have the same expressions as Eqs. (12), (13), and (14) of Ref. \cite{Wu:2017weo} for the cases $J=5/2$, $J=3/2$, and $J=1/2$, respectively. The base state is $[(uud)^S_{MA}(c\bar{c})^1_8]^{5/2}$ for the case $J=5/2$, where the superscript $S$ indicates that the spin of $uud$ is 3/2 and the subscript $MA$ means that the color representation of $uud$ is $8^{MA}$, i.e. the first two quarks are antisymmetric. The spins of the $c\bar{c}$ and the pentaquark as well as the color representation of $c\bar{c}$ are explicitly given. The base vectors for the cases $J=3/2$ and $J=1/2$ are
\begin{align}
J=\frac32:&\qquad ([(uud)^S_{MA}(c\bar{c})^1_8]^{\frac32}, [(uud)^S_{MA}(c\bar{c})^0_8]^{\frac32}, [(uud)^{MS}_{MA}(c\bar{c})^1_8]^{\frac32}, [(uud)^{MA}_{MS}(c\bar{c})^0_8]^{\frac32})^T; \nonumber\\
J=\frac12: &\qquad ([(uud)^S_{MA}(c\bar{c})^1_8]^{\frac12}, [(uud)^{MS}_{MA}(c\bar{c})^1_8]^{\frac12}, [(uud)^{MS}_{MA}(c\bar{c})^0_8]^{\frac12}, [(uud)^{MA}_{MS}(c\bar{c})^1_8]^{\frac12}, [(uud)^{MA}_{MS}(c\bar{c})^1_8]^{\frac12})^T.
\end{align}

When one considers the $q_1q_2q_3q_4\bar{q}_5=udsc\bar{c}$ states, the CMI matrices have the form \cite{Buccella:2006fn}
\begin{eqnarray}
\langle H_{CM}\rangle_J=\left(\begin{array}{cc}X&Y\\Y^T&Z\end{array}\right),
\end{eqnarray}
where $X$ and $Z$ are symmetric matrices. The base states for the $X$ part are obtained by replacing $uud$ with $uds$. Those for the $Z$ part are
\begin{align}
J=\frac52:&\qquad [(uds)^S_{MS}(c\bar{c})_8^1]^{\frac52};\nonumber\\
J=\frac32:&\qquad [(uds)^S_{MS}(c\bar{c})^1_8]^{\frac32}, [(uds)^S_{MS}(c\bar{c})^0_8]^{\frac32}, [(uds)^{MS}_{MS}(c\bar{c})^1_8]^{\frac32}, [(uds)^{MA}_{MA}(c\bar{c})^0_8]^{\frac32};\nonumber\\
J=\frac12:&\qquad [(uds)^S_{MS}(c\bar{c})^1_8]^{\frac12}, [(uds)^{MS}_{MS}(c\bar{c})^1_8]^{\frac12}, [(uds)^{MS}_{MS}(c\bar{c})^0_8]^{\frac12}, [(uds)^{MA}_{MA}(c\bar{c})^1_8]^{\frac12}, [(uds)^{MA}_{MA}(c\bar{c})^0_8]^{\frac12}.
\end{align}
The $X$ expressions are slightly different from those given in Eqs. (12), (13), and (14) of Ref. \cite{Wu:2017weo} for the cases $J=5/2$, $J=3/2$, and $J=1/2$, respectively. For $Z$, the expressions are slightly different from those given in Eqs. (15), (16), and (17) of Ref. \cite{Wu:2017weo} for the cases $J=5/2$, $J=3/2$, and $J=1/2$, respectively. They can be obtained with the replacements $C_{13}\to (C_{13}+C_{23})/2$, $C_{14}\to (C_{14}+C_{24})/2$, and $C_{15}\to (C_{15}+C_{25})/2$.

The $Y$ matrices account for the isospin breaking effects which are related with only $C_{ux}-C_{dx}$ with $x$ standing for $s$, $c$, or $\bar{c}$. We here give their expressions explicitly. For convenience, we define the variables $\theta=C_{13}-C_{23}$, $\tau=C_{14}-C_{24}$, and $\eta=C_{15}-C_{25}$. For the $J=5/2$ case, one has
\begin{eqnarray}
Y&=&\frac{1}{\sqrt3}\Big[3(C_{13}-C_{23})+(C_{14}-C_{24})-4(C_{15}-C_{25})\Big].
\end{eqnarray}
For the $J=3/2$ case, we have
\begin{eqnarray}
Y&=&\left(\begin{array}{cccc}
\frac{\sqrt3}{9}(9\theta-2\tau+8\eta)&\frac{\sqrt5}{3}(\tau+4\eta)&-\frac{\sqrt{15}}{9}(\tau-4\eta)&-\frac{\sqrt{15}}{9}(7\tau+2\eta)\\
\frac{\sqrt5}{3}(\tau+4\eta)&\sqrt{3}\theta&\frac13(\tau+4\eta)&\frac13(7\tau-2\eta)\\
-\frac{\sqrt{15}}{9}(\tau-4\eta)&\frac13(\tau+4\eta)&-\frac{2\sqrt3}{9}(9\theta-\tau+4\eta)&\frac{\sqrt3}{9}(3\theta-7\tau-2\eta)\\
-\frac{9\sqrt{15}}{9}(\tau+2\eta)&\frac53(\tau-2\eta)&-\frac{5\sqrt3}{9}(3\theta+\tau+2\eta)&0\end{array}\right).
\end{eqnarray}
This matrix is not symmetric. For the $J=1/2$ case, we have
\begin{eqnarray}
Y&=&\left(\begin{array}{ccccc}
\frac{\sqrt3}{9}(9\theta-5\tau+20\eta)&-\frac{\sqrt6}{9}(\tau-4\eta)&-\frac{\sqrt2}{3}(\tau+4\eta)&-\frac{\sqrt6}{9}(7\tau+2\eta)&-\frac{\sqrt2}{3}(7\tau-2\eta)\\
-\frac{\sqrt6}{9}(\tau-4\eta)&-\frac{2\sqrt3}{9}(9\theta+2\tau-8\eta)&\frac23(\tau+4\eta)&\frac{\sqrt3}{9}(3\theta+14\tau+4\eta)&-\frac13(7\tau-2\eta)\\
-\frac{\sqrt2}{3}(\tau+4\eta)&\frac23(\tau+4\eta)&-2\sqrt3\theta&-\frac13(7\tau-2\eta)&\frac{1}{\sqrt3}\theta\\
-\frac{5\sqrt6}{9}(\tau+2\eta)&-\frac{5\sqrt3}{9}(3\theta-2\tau-4\eta)&-\frac53(\tau-2\eta)&0&0\\
-\frac{5\sqrt2}{3}(\tau-2\eta)&-\frac53(\tau-2\eta)&-\frac{5}{\sqrt3}\theta&0&0\end{array}\right).
\end{eqnarray}

The pentaquark masses will be estimated with the formula \cite{Liu:2019zoy}
\begin{eqnarray}\label{massref}
M_{penta}=[M_{ref}-\langle H_{CM}\rangle_{ref}]+\langle H_{CM}\rangle_{penta},
\end{eqnarray}
where we take the reference mass scale $M_{ref}$ as the threshold of a $(qqc)$-$(q\bar{c})$ baryon-meson channel. Because the model does not involve dynamics and a definite prediction cannot be made, we have to choose the channel so that the theoretical calculations are close to the realistic case. Probably the adoption of the threshold resulting in high-mass pentaquarks is a reasonable selection \cite{Wu:2018xdi} and we use this strategy in the following parts. The method with Eq. \eqref{massref} has also been applied recently to the $cs\bar{c}\bar{s}$ \cite{Wu:2016gas}, $QQ\bar{Q}\bar{Q}$ \cite{Wu:2016vtq}, $QQ\bar{Q}\bar{q}$ \cite{Chen:2016ont}, $qq\bar{Q}\bar{Q}$ \cite{Luo:2017eub}, $QQqq\bar{q}$ \cite{Zhou:2018pcv}, $QQQq\bar{q}$ \cite{Li:2018vhp}, and $qqQQ\bar{Q}$ \cite{An:2019idk} systems.

We have considered the mixing effects between different color-spin structures of pentaquark states. The effective color-magnetic interactions between a pair of quark components are different for states with different quantum numbers. If we define a measure to reflect the interaction strength, the value of the CMI for a pentaquark state can be written as
\begin{eqnarray}
\langle H_{CM}\rangle_{penta}=\sum_{i<j}K_{ij}C_{ij},
\end{eqnarray}
where $K_{ij}$ is the defined measure \cite{Liu:2019zoy,Li:2018vhp}. In the color-magnetic model, a coupling parameter $C_{ij}$ for one state should, in principle, differ from the same parameter for another state, since it relies on the spacial wave function. In practice, they are extracted from conventional meson and baryons and are treated as universal parameters. It is obvious that the adopted parameters result in uncertainties in mass splittings for pentaquark states. The measure $K_{ij}$ provides a method to reflect the effects on the change of coupling parameters.

\subsection{Rearrangement decays}

To explore the decay properties of the pentaquark states through rearrangement mechanisms, we here use a very simple scheme to estimate the amplitude ${\cal M}$: the quark-level Hamiltonian for decay is taken as a constant $H_{decay}=\alpha$. The two-body partial decay widths are then calculated with the standard formula,
\begin{eqnarray}
\Gamma=|{\cal M}|^2\frac{|\vec{p}_1|}{8\pi M_{penta}^2},
\end{eqnarray}
where $|\vec{p}_1|$ is the three-momentum of a final state in the center-of-mass frame. The amplitude squared $|{\cal M}|^2$ is then calculated by finding the possibility of the final state existing in the initial state. Because the Pauli principle has effects on the identical $u$ quarks in the initial state wave function while it has no effect on quarks in the wave function of the final state $(udc)$-$(u\bar{c})$, perhaps a question may arise in re-coupling the initial wave function. One may calculate $|{\cal M}|^2$ by re-coupling the final wave function into the form of the above pentaquark base states, i.e.
\begin{eqnarray}
(q_1q_2c)(q_3\bar{c})=\sum_i Y_i(q_1q_2q_3c\bar{c})_i, \quad or\quad \sum_i Y_i(q_1q_3q_2c\bar{c})_i, \quad or \quad \sum_i Y_i(q_3q_1q_2c\bar{c})_i,
\end{eqnarray}
where $Y_i$'s are obtained from the $SU(3)$ and $SU(2)$ Clebsch-Gordan coefficients. An initial state wave function can also be expressed in a similar form
\begin{eqnarray}
\psi_{penta}=\sum_i X_i(q_1q_2q_3c\bar{c})_i.
\end{eqnarray}
Then $|{\cal M}|^2=\alpha^2|\sum_i X_iY_i|^2$. In the pentaquark picture we consider, this simple estimation method cannot describe the hidden-charm decays.

When comparing numerical results with experimental results, we will use the ratios of total widths between the $P_c$ states,
\begin{eqnarray}\label{expratios}
\Gamma(P_c(4440)^+):\Gamma(P_c(4457)^+)=3.2^{+2.1}_{-3.5}:1,\nonumber\\
\Gamma(P_c(4440)^+):\Gamma(P_c(4312)^+)=2.1^{+1.5}_{-1.5}:1,\nonumber\\
\Gamma(P_c(4312)^+):\Gamma(P_c(4457)^+)=1.5^{+1.0}_{-1.7}:1.
\end{eqnarray}

\section{The $uudc\bar{c}$ pentaquark states}\label{sec3}

\begin{table}[htbp]
\caption{The results for the $(uud)_{8_c}(c\bar{c})_{8_c}$ pentaquark states in the case $m_u\neq m_d$. The CMI eigenvalues in the second column and the pentaquark masses estimated with the $\Sigma_c^{++}D^-$ threshold in the third column are given in units of MeV.
}\label{massKij-uud}
\begin{tabular}{c|cc|ccccccc}\hline
$J^{P}$ &Eigenvalue &$\Sigma_c^{++}D^-$&$K_{uu}$&$K_{ud}$&$K_{uc}$&$K_{u\bar{c}}$&$K_{dc}$&$K_{d\bar{c}}$&$K_{c\bar{c}}$\\\hline
$\frac52^{-}$ &100.9&4519.6&2.67&-0.67&4.67&1.33&1.33&4.67&-0.67\\
$\frac32^{-}$ &182.2&4600.9&3.34&6.66&2.21&-2.18&1.13&-1.16&-0.67\\
&77.6&4496.3&2.82&-1.78&3.03&3.05&-0.37&2.60&1.35\\
&29.8&4448.4&2.81&-1.70&-8.71&2.86&0.14&3.55&-0.37\\
&-79.3&4339.4&3.03&-3.18&3.47&-3.74&0.44&-10.32&-0.31\\

$\frac12^{-}$&268.3&4687.0&3.34&6.66&1.56&6.13&0.79&3.06&0.63\\
&145.0&4563.7&3.34&6.66&-5.99&-1.65&-3.03&-0.88&0.70\\
&32.4&4451.0&3.09&-3.66&-0.43&4.82&-2.72&2.49&0.92\\
&-79.5&4339.2&3.30&-5.14&-2.02&-8.79&0.12&3.09&-0.22\\
&-129.1&4289.6&2.93&-2.52&-7.12&-4.52&-1.82&-11.09&-0.03\\
\hline
\end{tabular}
\end{table}

\begin{figure}[htbp]
\includegraphics[width=230pt]{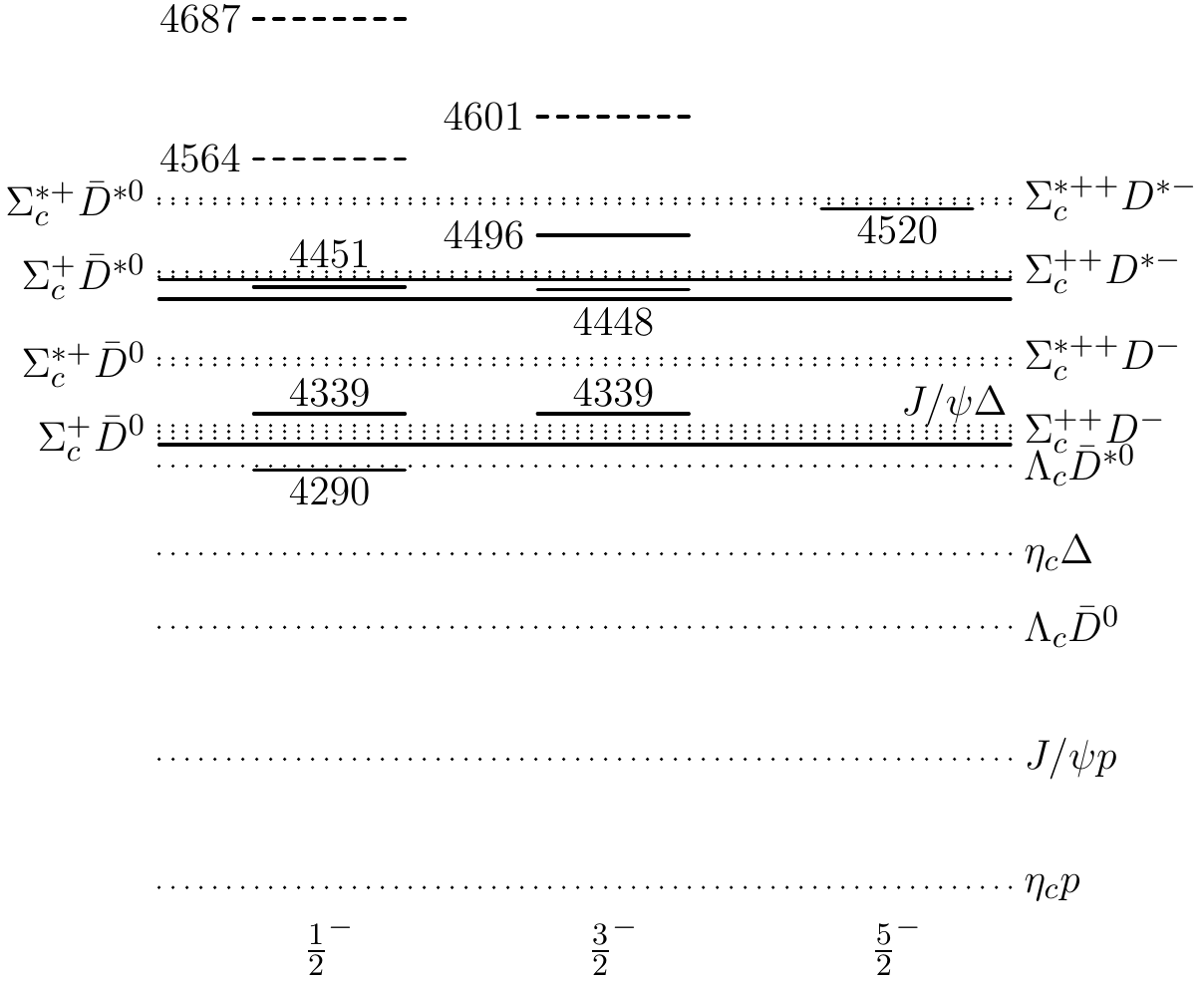}
\caption{Relative positions for the obtained $(uud)_{8_c}(c\bar{c})_{8_c}$ pentaquark states (solid and dashed short lines), relevant meson-baryon thresholds (dotted long lines), and the newly observed three $P_c$ states (solid long lines). The pentaquark masses are given in units of MeV. We have used dashed short lines to denote pentaquarks having dominantly $I=3/2$ components.}\label{fig-uud}
\end{figure}

With the parameters determined in Table \ref{parameter}, we get the numerical CMI matrices for the $uudc\bar{c}$ case. Then their eigenvalues, the pentaquark masses estimated with the $\Sigma_c^{++}D^-$ threshold, and the measures $K_{ij}$'s can be obtained. These results are shown in Table \ref{massKij-uud}. Fig. \ref{fig-uud} illustrates the relative positions of the pentaquark states. From the CMI eigenvectors, it is clear that the highest three states, one with $J=3/2$ and two with $J=1/2$, have dominantly $I=3/2$ components. The isospin breaking effects in pentaquark spectrum are not significant, which is observed from the masses differences $|M_{dduc\bar{c}}-M_{uudc\bar{c}}|<3$ MeV. From the obtained $K_{ij}$'s, one may check whether the effective CMI is attractive or repulsive and understand roughly the effects on pentaquark masses due to the change of coupling parameters.

From Fig. \ref{fig-uud}, the $J^P=1/2^-$ state with mass 4451 MeV and the $J^P=3/2^-$ state with mass 4448 MeV agree nicely with the observed $P_c(4457)^+$ and $P_c(4440)^+$, respectively. This fact indicates the reasonability in estimating the masses of $qqqc\bar{c}$ pentaquarks with Eq. \eqref{massref}. However, Ali and Parkhomenko stressed in Ref. \cite{Ali:2019npk} that only the mass closeness cannot support the molecule interpretation for the $P_c$ states. Similarly, we cannot assign the $P_c(4457)^+$ to be a $J^P=1/2^-$ $(uud)_{8_c}(c\bar{c})_{8_c}$ pentaquark and the $P_c(4440)^+$ to be a $J^P=3/2^-$ $(uud)_{8_c}(c\bar{c})_{8_c}$ pentaquark just from Fig. \ref{fig-uud}. Both the mass uncertainties and decay properties should be considered.

There are two sources contributing to the mass uncertainties in the present estimation method, the reference scale and the values of coupling parameters. We temporarily assume that the used reference scale is reasonable and its effects will be discussed later. Then the masses are affected mainly by the uncertainties of $C_{ij}$'s in Eq. \eqref{Hamiltonian}. It is still an open question whether the coupling constants derived from the conventional hadrons are applicable to multiquark states or not. In the case that the $C_{ij}$'s for multiquark states are not far from those for conventional hadrons, one finds that it is possible to reverse the mass order by adjusting the values of coupling parameters. Then assigning the $P_c(4457)^+$ to be a $J^P=3/2^-$ state and the $P_c(4440)^+$ to be a $J^P=1/2^-$ state is also possible. The mass uncertainties caused by the adjustment $C_{ij}\to C_{ij}\pm 1$ MeV can reach tens of MeV, which may be understood with the $K_{ij}$'s shown in Table \ref{massKij-uud}. In the following discussions, we always assume that the obtained CMI eigenvalues in Table \ref{massKij-uud} are consistent with experiments and are thus acceptable.

From Fig. \ref{fig-uud}, there are three pentaquark states around 4.3 GeV and each of them can be assigned as the $P_c(4312)^+$. Then the possible $J^P$ assignments for the three $P_c$ states are: $(P_c(4457)^+,P_c(4440)^+)\to (1/2^-,3/2^-)$ or $(3/2^-,1/2^-)$ while $P_c(4312)^+\to1/2^-$ or $3/2^-$.  To get more information about their $J^P$ assignments, one needs to study the decay properties of the pentaquarks.

\begin{table}[htbp]
\caption{Rearrangement decays into open-charm channels for the $(uud)_{8_c}(c\bar{c})_{8_c}$ pentaquark states in the case $m_u\neq m_d$. The numbers in the parentheses are ($100 |{\cal M}|^2/\alpha^2$, $10^7\Gamma/\alpha^2\cdot$MeV). The symbol ``$-$'' means that the decay is forbidden.}\label{decay-uud-compact}
\begin{tabular}{c|ccccccccc}\hline
States&\multicolumn{8}{c}{Channels}&$\Gamma_{sum}$\\\hline
$J^P=\frac52^-$&$\Sigma_c^{*+}\bar{D}^{*0}$&$\Sigma_c^{*++}D^{*-}$&&&&&&&\\
4519.6&(22.2,$-$)&(88.9,$-$)&&&&&&&0.0\\
\hline
$J^P=\frac32^-$&$\Lambda_c^+\bar{D}^{*0}$&$\Sigma_c^{*+}\bar{D}^0$&$\Sigma_c^{*++}D^-$&$\Sigma_c^+\bar{D}^{*0}$&$\Sigma_c^{++}D^{*-}$&$\Sigma_c^{*+}\bar{D}^{*0}$&$\Sigma_c^{*++}D^{*-}$&&\\
4600.9&(0.0,0.0)&(14.8,1.9)&(14.9,1.9)&(5.0,0.5)&(4.9,0.5)&(24.9,1.9)&(24.2,1.8)&&8.7\\
4496.3&(8.0,1.0)&(1.8,0.2)&(6.9,0.7)&(2.5,0.1)&(9.8,0.5)&(15.1,$-$)&(61.8,$-$)&&2.5\\
4448.4&(7.4,0.9)&(0.7,0.1)&(2.5,0.2)&(18.4,$-$)&(74.0,$-$)&(0.6,$-$)&(2.6,$-$)&&1.1\\
4339.4&(18.0,1.2)&(16.1,$-$)&(64.5,$-$)&(0.0,$-$)&(0.1,$-$)&(0.1,$-$)&(0.3,$-$)&&1.2\\
\hline
$J^P=\frac12^-$&$\Lambda_c^+\bar{D}^0$&$\Lambda_c^+\bar{D}^{*0}$&$\Sigma_c^+\bar{D}^0$&$\Sigma_c^{++}D^-$&$\Sigma_c^+\bar{D}^{*0}$&$\Sigma_c^{++}D^{*-}$&$\Sigma_c^{*+}\bar{D}^{*0}$&$\Sigma_c^{*++}D^{*-}$\\
4687.0&(0.0,0.0)&(0.0,0.0)&(0.9,0.1)&(0.9,0.1)&(7.0,0.9)&(6.9,0.9)&(36.7,4.0)&(36.4,4.0)&10.1\\
4563.7&(0.0,0.0)&(0.0,0.0)&(13.9,1.9)&(14.0,1.9)&(27.8,2.6)&(27.3,2.5)&(2.9,0.2)&(2.8,0.2)&9.2\\
4451.0&(1.3,0.2)&(20.1,2.4)&(0.7,0.1)&(2.5,0.3)&(7.1,$-$)&(29.0,$-$)&(7.2,$-$)&(29.2,$-$)&2.9\\
4339.2&(31.7,4.2)&(0.3,0.0)&(0.4,0.0)&(1.4,0.1)&(6.3,$-$)&(25.6,$-$)&(4.8,$-$)&(19.4,$-$)&4.3\\
4289.6&(0.3,0.0)&(12.9,$-$)&(17.5,$-$)&(70.1,$-$)&(0.0,$-$)&(0.1,$-$)&(0.3,$-$)&(1.1,$-$)&0.0\\
\hline
\end{tabular}
\end{table}

In Table \ref{decay-uud-compact}, we present the results of decay widths for the rearrangement channels. There is no hidden-charm decay mode in the present simple model, which is a feature consistent with the analysis given in Ref. \cite{Cao:2019kst}, ${\cal B}(P_c\to J/\psi p)<2\%$. If we assume the approximation $\Gamma_{tot}\approx \Gamma_{sum}$ for the pentaquark states, one gets the ratio
\begin{eqnarray}
\frac{\Gamma_{tot}(P_c(4451))}{\Gamma_{tot}(P_c(4448))}\approx 2.6,
\end{eqnarray}
which is consistent with the ratio between $P_c(4440)^+$ and $P_c(4457)^+$. Then we have a possible assignment in the present pentaquark picture: the $P_c(4457)^+$ and $P_c(4440)^+$ have quantum numbers $J^P=3/2^-$ and $1/2^-$, respectively. Unfortunately, one cannot interpret the $P_c(4312)^+$ as the lowest $(uud)_{8_c}(c\bar{c})_{8_c}$ with $J^P=1/2^-$. This lowest state has a very narrow width because of the small coupling with the $\Lambda_c\bar{D}^0$ channel. If we add 22.3 MeV to its mass, this state can also decay into $\Lambda_c\bar{D}^{*0}$ and the $\Gamma_{sum}$ in Table \ref{decay-uud-compact} is 0.6 now. We still cannot get the ratio $\Gamma(P_c(4440)^+):\Gamma(P_c(4312)^+)$ consistent with the LHCb experiment. However, one may interpret the $P_c(4312)^+$ as the lowest $J^P=3/2^-$ state by noting
\begin{eqnarray}
\frac{\Gamma_{tot}(P_c(4451))}{\Gamma_{tot}(P_c(4339))}\approx 2.4, \quad \frac{\Gamma_{tot}(P_c(4339))}{\Gamma_{tot}(P_c(4448))}\approx1.1.
\end{eqnarray}
The former is consistent with the ratio between $P_c(4440)^+$ and $P_c(4312)^+$ and the latter is consistent with that between $P_c(4312)^+$ and $P_c(4457)^+$. The order of decay widths with this assignment is also consistent with the observed one, $\Gamma(P_c(4440)^+)>\Gamma(P_c(4312)^+)>\Gamma(P_c(4457)^+)$. If the above assignments are correct, probably three states should exist around 4312 MeV, not just one. According to Table \ref{decay-uud-compact}, the broad one of the other two $J^P=1/2^-$ states has a width around 30 MeV and the narrow one has a width less than 1 MeV.

Let us check predictions with the above assignments further, i.e. the $P_c(4457)^+$ ($P_c(4312)^+$) corresponds to the $J^P=3/2^-$ pentaquark with mass 4448 (4339) MeV and the $P_c(4440)^+$ corresponds to the $J^P=1/2^-$ state with mass 4451 MeV in Table \ref{decay-uud-compact}. For the $P_c(4312)^+$ state, it mainly decays into $\Lambda_c^+\bar{D}^{*0}$. For the $P_c(4457)^+$, its dominant decay channel is also $\Lambda_c^+\bar{D}^{*0}$. The partial width is larger than that into the $\Sigma_c^*\bar{D}$ mode. If the isospin symmetry breaking is not considered, the ratio $\Gamma(P_c(4457)^+\to \Sigma_c^{*++}D^-):\Gamma(P_c(4457)^+\to \Sigma_c^{*+}\bar{D}^0)=|{\cal M}(P_c(4457)^+\to \Sigma_c^{*++}D^-)|^2:{\cal M}(P_c(4457)^+\to \Sigma_c^{*+}\bar{D}^0)|^2=4:1$ should be satisfied strictly in the model we use. This value is from the isospin wave function. Since we consider the isospin breaking effects, this ratio is slightly smaller ($\sim3.9$). For the $P_c(4440)^+$, it may decay into $\Sigma_c\bar{D}$ channels, not $\Sigma_c^*\bar{D}$. Large ratio $\Gamma(P_c(4440)^+\to \Sigma_c^{++}D^-):\Gamma(P_c(4440)^+\to \Sigma_c^{+}\bar{D}^0)\approx 4:1$ is also expected.

\begin{table}[!h]
\caption{Rearrangement decays into open-charm channels for the $P_c(4457)^+$, $P_c(4440)^+$, and $P_c(4312)^+$ by assigning them as $J^P=\frac32^-$, $\frac12^-$, and $\frac32^-$ $(uud)_{8_c}(c\bar{c})_{8_c}$ pentaquark states, respectively, in the case $m_u\neq m_d$. The numbers in the parentheses are ($100 |{\cal M}|^2/\alpha^2$, $10^7\Gamma/\alpha^2\cdot$MeV). The symbol ``$-$'' means that the decay is forbidden.}\label{decay-uud-compact-exp}
\begin{tabular}{c|ccccccccc}\hline
States&\multicolumn{8}{c}{Channels}&$\Gamma_{sum}$\\\hline
$J^P=\frac32^-$&$\Lambda_c^+\bar{D}^{*0}$&$\Sigma_c^{*+}\bar{D}^0$&$\Sigma_c^{*++}D^-$&$\Sigma_c^+\bar{D}^{*0}$&$\Sigma_c^{++}D^{*-}$&$\Sigma_c^{*+}\bar{D}^{*0}$&$\Sigma_c^{*++}D^{*-}$\\
4457.3&(7.4,0.9)&(0.7,0.1)&(2.5,0.2)&(18.4,$-$)&(74.0,$-$)&(0.6,$-$)&(2.6,$-$)&&1.1\\
4311.9&(18.0,0.8)&(16.1,$-$)&(64.5,$-$)&(0.0,$-$)&(0.1,$-$)&(0.1,$-$)&(0.3,$-$)&&0.8\\
$J^P=\frac12^-$&$\Lambda_c^+\bar{D}^0$&$\Lambda_c^+\bar{D}^{*0}$&$\Sigma_c^+\bar{D}^0$&$\Sigma_c^{++}D^-$&$\Sigma_c^+\bar{D}^{*0}$&$\Sigma_c^{++}D^{*-}$&$\Sigma_c^{*+}\bar{D}^{*0}$&$\Sigma_c^{*++}D^{*-}$\\
4440.3&(1.3,0.2)&(20.1,2.3)&(0.7,0.1)&(2.5,0.3)&(7.1,$-$)&(29.0,$-$)&(7.2,$-$)&(29.2,$-$)&2.8\\
\hline
\hline
\end{tabular}
\end{table}

Now we consider the effects by choosing different reference scales. If we add 22.3 MeV to all the pentaquark masses in Table \ref{massKij-uud}, i.e. the lowest $J^P=1/2^-$ pentaquark has the mass of the observed $P_c(4312)^+$, the $\Gamma_{sum}$'s for our pentaquarks assigned as $3/2^-$ $P_c(4457)^+$, $1/2^-$ $P_c(4440)^+$, and $3/2^-$ $P_c(4312)^+$ become 3.5, 4.1, and 1.4, respectively. The above assignments are still allowed once the LHCb errors are included. However, the ratio $\Gamma(P_c(4440)^+):\Gamma(P_c(4312)^+)$ still cannot be explained if the $P_c(4312)^+$ is treated as the lowest $J^P=1/2^-$ pentaquark state. If we add -27.5 MeV to all the pentaquark masses in Table \ref{massKij-uud}, i.e. the lowest $J^P=3/2^-$ pentaquark has the mass of the observed $P_c(4312)^+$, the above assignments are not affected much. If we use different scales for the three states, i.e. adjust the masses to their experimental values, we get results shown in Table \ref{decay-uud-compact-exp}. The $\Gamma_{sum}$'s are close to those obtained by adding -27.5 MeV to all the pentaquark masses in Table \ref{massKij-uud}.

From the above discussions, one may assign the $P_c(4457)^+$ as a $J^P=3/2^-$ pentaquark, $P_c(4440)^+$ as a $J^P=1/2^-$ pentaquark, and $P_c(4312)^+$ as a $J^P=3/2^-$ pentaquark in the $(uud)_{8_c}(c\bar{c})_{8_c}$ configuration. Predictions for other pentaquark states can be found in Fig. \ref{fig-uud} and Table \ref{decay-uud-compact}. Further analyses about the $J/\psi p$ spectrum and open-charm decay channels can answer whether the proposed assignments are correct or not.

\begin{table}[htbp]
\caption{Rearrangement decays for the $S$-wave $(nnc)(n\bar{c})$ ($n=u,d$) molecular states in the isospin-symmetric case. The numbers in the parentheses are ($100 |{\cal M}|^2/{\alpha^\prime}^2$, $10^7\Gamma/{\alpha^\prime}^2\cdot$MeV). The symbol ``$-$'' means that the decay is forbidden. We do not show open-charm decay channels since their couplings with the decay channels vanish.}\label{decay-uud-mole}
\begin{tabular}{c|cccc}\hline
States &\multicolumn{3}{c}{Channels}&$\Gamma_{sum}$\\\hline
$J^P=\frac52^-$&$\Delta^{+}J/\psi$&&&\\
$(\Sigma_{c}^{*}\bar{D}^{*})^{I=\frac{3}{2}}$&(11.1,1.3)&&&1.3\\
$(\Sigma_{c}^{*}\bar{D}^{*})^{I=\frac{1}{2}}$&(0.0,$-$)&&&0.0\\
\hline
$J^P=\frac32^-$&$pJ/\psi$&$\Delta^{+}\eta_{c}$&$\Delta^{+}J/\psi$&\\
$(\Sigma_{c}^{*}\bar{D}^{*})^{I=\frac{3}{2}}$&(0.0,$-$)&(4.6,0.7)&(0.3,0.0)&0.7\\
$(\Sigma_{c}^{*}\bar{D}^{*})^{I=\frac{1}{2}}$&(6.2,1.1)&(0.0,$-$)&(0.0,$-$)&1.1\\
$(\Sigma_{c}\bar{D}^{*})^{I=\frac{3}{2}}$&(0.0,$-$)&(3.7,0.5)&(6.2,0.6)&1.1\\
$(\Sigma_{c}\bar{D}^{*})^{I=\frac{1}{2}}$&(1.2,0.2)&(0.0,$-$)&(0.0,$-$)&0.2\\
$(\Sigma_{c}^{*}\bar{D})^{I=\frac{3}{2}}$&(0.0,$-$)&(2.8,0.3)&(4.6,0.3)&0.6\\
$(\Sigma_{c}^{*}\bar{D})^{I=\frac{1}{2}}$&(3.7,0.6)&(0.0,$-$)&(0.0,$-$)&0.6\\
$(\Lambda_{c}\bar{D}^{*})^{I=\frac{1}{2}}$&(0.0,$-$)&(0.0,$-$)&(0.0,$-$)&0.0\\
\hline
$J^P=\frac12^-$&$p\eta_{c}$&$pJ/\psi$&$\Delta^{+}J/\psi$\\
$(\Sigma_{c}^{*}\bar{D}^{*})^{I=\frac{3}{2}}$&(0.0,$-$)&(0.0,$-$)&(1.2,0.1)&0.1\\
$(\Sigma_{c}^{*}\bar{D}^{*})^{I=\frac{1}{2}}$&(7.4,1.5)&(2.5,0.4)&(0.0,$-$)&1.9\\
$(\Sigma_{c}\bar{D}^{*})^{I=\frac{3}{2}}$&(0.0,$-$)&(0.0,$-$)&(2.5,0.2)&0.2\\
$(\Sigma_{c}\bar{D}^{*})^{I=\frac{1}{2}}$&(0.9,0.2)&(7.7,1.3)&(0.0,$-$)&1.5\\
$(\Sigma_{c}\bar{D})^{I=\frac{3}{2}}$&(0.0,$-$)&(0.0,$-$)&(7.4,$-$)&0.0\\
$(\Sigma_{c}\bar{D})^{I=\frac{1}{2}}$&(2.8,0.5)&(0.9,0.1)&(0.0,$-$)&0.6\\
$(\Lambda_{c}\bar{D}^{*})^{I=\frac{1}{2}}$&(0.0,$-$)&(0.0,$-$)&(0.0,$-$)&0.0\\
$(\Lambda_{c}\bar{D})^{I=\frac{1}{2}}$&(0.0,$-$)&(0.0,$-$)&(0.0,$-$)&0.0\\\hline
\end{tabular}
\end{table}

In the molecule picture, the masses of the observed $P_c$ states are easy to understand. The decay width of $P_c(4312)^+$ can also be understood \cite{Shen:2017ayv}. Now we concentrate on their decay properties within the simple rearrangement model, $H_{decay}=\alpha^\prime$. Note that the value of $\alpha^\prime$ in this case may be different from $\alpha$ in the compact pentaquark case. As a simple comparison study, we just consider the $S$-wave molecules in the isospin-symmetric case with the approximation $M_{mole}\approx M_{threshold}$. In Table \ref{decay-uud-mole}, we present the calculated partial decay widths for various states. The adopted simple model cannot give nonvanishing decay widths for the open-charm channels. This probably indicates that the hidden-charm decays are important in the molecule picture or that the decay model is oversimplified. For higher charmonium states, e.g. $\psi(3770)$, their open-charm channels should dominate the decays. If this is also the case for hidden-charm pentaquarks, the estimation of the total decay widths relies on the branching ratios to $J/\psi p$. Here, we use the constraints obtained in Ref. \cite{Cao:2019kst},
\begin{eqnarray}
1.2\% > {\cal B}(P_c(4312)^+\to J/\psi p)> 0.05\%,\nonumber\\
2\% > {\cal B}(P_c(4440)^+\to J/\psi p)> 0.2\%,\nonumber\\
2\% > {\cal B}(P_c(4457)^+\to J/\psi p)> 0.1\%,
\end{eqnarray}
which indicate that the branching ratios into $J/\psi p$ for the $P_c(4457)^+$ and $P_c(4440)^+$ states have similar values while that of $P_c(4312)^+$ is smaller. From Fig. \ref{fig-uud} and Table \ref{decay-uud-mole}, the correspondence between molecules and the three $P_c$ states would be: the $P_c(4312)^+$ is a $J^P=1/2^-$ $\Sigma_c\bar{D}$ state and the $P_c(4440)^+$ and $P_c(4457)^+$ are two $\Sigma_c\bar{D}^*$ states with different angular momenta. If $J^P$ of $P_c(4440)^+$ are $1/2^-$, we may use, for example, ${\cal B}(P_c(4440)^+\to J/\psi p)\sim2\%$, ${\cal B}(P_c(4457)^+\to J/\psi p)\sim 1\%$, and ${\cal B}(P_c(4312)^+\to J/\psi p)\sim0.2\%$ to understand the ratios in Eq. \eqref{expratios}. If $J^P$ of $P_c(4440)^+$ are $3/2^-$, we need a small ${\cal B}(P_c(4440)^+\to J/\psi p)$ and a large ${\cal B}(P_c(4457)^+\to J/\psi p)$ to understand $\Gamma_{tot}(P_c(4440)^+)>\Gamma_{tot}(P_c(4457)^+)$. The former assignment seems to be favored, but the latter assignment is also possible. Then more experimental data are needed to confirm the $J^P$ assignments in the molecule picture.

From the $J^P$ assignments for the three $P_c$ states in the pentaquark picture we consider and the molecule picture, both pictures can explain the masses and decay properties for the $P_c(4457)^+$ and $P_c(4440)^+$ states consistently. For the $P_c(4312)^+$, further measurements will be helpful to understand its nature, a $1/2^-$ state or a $3/2^-$ state. However, detailed coupled-channel investigations in the molecule picture are definitely needed in order to draw a further conclusion on the theoretical side. On the experimental side, searching for more states around the $P_c(4312)^+$ and measuring their properties are also strongly called for.

\section{The $udsc\bar{c}$ pentaquark states}\label{sec4}

\begin{table}[htbp]
\caption{The results for the $(uds)_{8_c}(c\bar{c})_{8_c}$ pentaquark states in the case $m_u\neq m_d$. The CMI eigenvalues in the second column and the pentaquark masses estimated with the $\Xi_c^{\prime+}D^-$ threshold in the third column are given in units of MeV.}\label{massKij-uds}
\begin{tabular}{c|cc|cccccccccc}\hline
$J^{P}$ & Eigenvalue &$\Xi_c^{\prime+}D^-$&$K_{ud}$&$K_{us}$&$K_{uc}$&$K_{u\bar{c}}$&$K_{ds}$&$K_{dc}$&$K_{d\bar{c}}$&$K_{sc}$&$K_{s\bar{c}}$&$K_{c\bar{c}}$\\\hline
$\frac52^{-}$ &104.6&4670.1&2.67&-0.31&2.34&0.63&-0.36&2.32&0.70&1.33&4.67&-0.67\\
&75.8&4641.2&-1.33&1.64&1.66&3.37&1.69&1.68&3.30&2.67&-0.67&-0.67\\

$\frac32^{-}$ &132.2&4697.7&3.43&3.25&0.97&-0.87&3.25&0.98&-0.94&1.39&-1.45&-0.66\\
&90.5&4655.9&2.87&-1.11&1.96&1.76&-1.27&1.92&1.75&-0.16&3.15&0.87\\
&54.3&4619.7&-2.04&1.38&0.30&2.30&1.51&0.34&2.28&2.37&1.30&1.24\\
&39.6&4605.1&2.79&-0.73&-4.86&1.14&-0.83&-4.76&1.26&-0.13&4.17&0.05\\
&2.7&4568.1&-1.86&1.43&-1.21&3.01&1.54&-1.32&2.97&-6.45&0.53&-0.28\\
&-62.5&4503.0&2.92&-1.10&1.94&-1.75&-1.45&1.85&-2.38&0.24&-11.20&-0.26\\
&-97.4&4468.0&-1.83&0.19&0.76&-6.67&0.55&0.85&-6.04&2.14&-0.20&-0.31\\
&-159.5&4405.9&-6.27&-3.31&0.14&1.07&-3.28&0.15&1.08&0.60&3.70&-0.65\\

$\frac12^{-}$ &218.2&4783.7&3.36&3.32&0.74&3.09&3.31&0.75&3.08&0.89&2.99&0.64\\
&93.1&4658.6&3.41&3.27&-2.82&-0.67&3.27&-2.83&-0.72&-3.38&-1.07&0.69\\
&56.6&4622.0&3.08&-1.86&-0.04&2.22&-2.10&-0.10&2.21&-2.32&3.01&1.16\\
&11.9&4577.3&-2.91&0.92&-2.15&2.60&1.13&-2.05&2.64&1.29&2.15&1.00\\
&-44.4&4521.0&3.29&-2.52&-0.95&-4.40&-2.67&-0.93&-4.21&-0.26&3.29&-0.35\\
&-100.0&4465.4&-3.92&0.86&-0.50&0.12&1.01&-0.51&-0.06&-0.85&-5.69&-0.29\\
&-117.5&4447.9&2.85&-0.79&-3.99&-1.88&-1.24&-3.83&-2.73&-1.60&-11.54&-0.14\\
&-140.1&4425.3&-1.89&-2.79&-0.95&-2.07&-2.33&-1.10&-1.29&-3.12&-0.25&0.66\\
&-168.2&4397.2&-5.95&-0.13&-1.35&-4.45&-0.10&-1.41&-4.35&-2.49&3.70&0.76\\
&-350.9&4214.6&-5.32&-4.28&0.01&-6.57&-4.27&0.00&-6.58&-0.16&-8.60&-0.13\\
\hline
\end{tabular}
\end{table}

In previous studies of pentaquark states in the $(qqq)_{8_c}(c\bar{c})_{8_c}$ configuration \cite{Wu:2017weo,Irie:2017qai}, a low mass $\Lambda$-type hidden-charm pentaquark state was found. The mass is roughly consistent with that obtained in the molecule picture \cite{Wu:2010jy}. From the above discussions, a $uudc\bar{c}$ state can be understood within both the compact pentaquark picture and the molecule picture. Probably similar feature also exists in the $udsc\bar{c}$ case once such a pentaquark state were observed. In this section, we discuss the properties of the $udsc\bar{c}$ states in the compact pentaquark picture. The decay properties of $S$-wave molecules are also given.

\begin{figure}
\includegraphics[width=240pt]{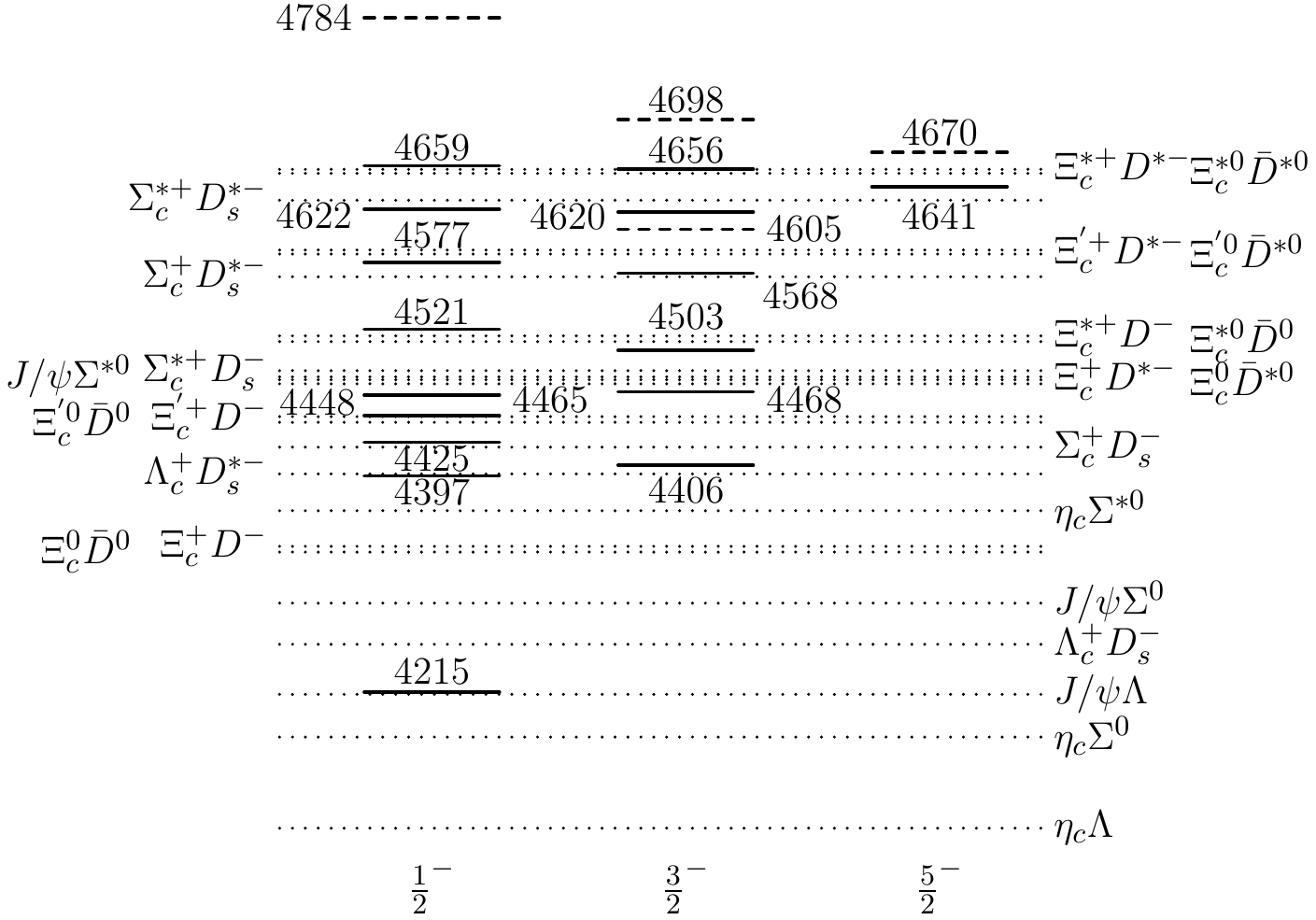}
\caption{Relative positions for the obtained $(uds)_{8_c}(c\bar{c})_{8_c}$ pentaquark states (solid and dashed short lines) and relevant meson-baryon thresholds (dotted long lines). The pentaquark masses are given in units of MeV. We have used dashed short lines to denote pentaquarks having dominantly $I=3/2$ components.}\label{fig-uds}
\end{figure}

Table \ref{massKij-uds} lists our results for the $(uds)_{8_c}(c\bar{c})_{8_c}$ pentaquark states in the case $m_u\neq m_d$, which includes the CMI eigenvalues, pentaquark masses, and measures of effective interactions. The pentaquark masses are estimated with the $\Xi_c^{\prime+}D^-$ threshold. Two states sensitive to the value of $C_{s\bar{c}}$ are observed. In Fig. \ref{fig-uds}, we plot the relative positions for such states as well as relevant baryon-meson thresholds. From the above discussions, the properties of the observed $P_c$ states can be understood with our estimations for the $(uud)_{8_c}(c\bar{c})_{8_c}$ pentaquark states. In a natural manner, we may assume that the obtained results in this section can explain the properties of some observed states in future measurements.

\begin{sidewaystable}\centering
\caption{Rearrangement decays into open-charm channels for the $(uds)_{8_c}(c\bar{c})_{8_c}$ pentaquark states in the case $m_u\neq m_d$. The numbers in the parentheses are ($100 |{\cal M}|^2/\alpha^2$, $10^7\Gamma/\alpha^2\cdot$MeV). The symbol ``$-$'' means that the decay is forbidden.}\label{decay-uds-compact}
\begin{tabular}{c|cccccccccccccccc}\hline
$J=5/2$&$\Sigma_c^{*+}D_s^{*-}$&$\Xi_c^{*0}\bar{D}^{*0}$&$\Xi_c^{*+}D^{*-}$&&&&&&&&&&&&&$\Gamma_{sum}$\\
4670.1&(88.9,4.9)&(21.6,0.8)&(22.8,0.8)&&&&&&&&&&&&&6.5\\
4641.2&(0.0,0.0)&(67.3,$-$)&(66.1,$-$)&&&&&&&&&&&&&0.0\\
\hline
$J=3/2$&$\Lambda_c^+D_s^{*-}$&$\Xi_c^0\bar{D}^{*0}$&$\Xi_c^+D^{*-}$&$\Sigma_c^{*+}D_s^-$&$\Xi_c^{*0}\bar{D}^0$&$\Xi_c^{*+}D^-$&$\Sigma_c^+D_s^{*-}$&$\Xi_c^{\prime0}\bar{D}^{*0}$&$\Xi_c^{\prime+}D^{*-}$&$\Sigma_c^{*+}D_s^{*-}$&$\Xi_c^{*0}\bar{D}^{*0}$&$\Xi_c^{*+}D^{*-}$&&&&$\Gamma_{sum}$\\
4697.7&(0.0,0.0)&(0.2,0.0)&(0.2,0.0)&(14.6,1.8)&(14.6,1.7)&(14.9,1.7)&(4.9,0.5)&(4.9,0.4)&(4.9,0.4)&(18.7,1.3)&(28.0,1.6)&(27.5,1.5)&&&&11.2\\
4655.9&(0.0,0.0)&(13.4,1.6)&(12.7,1.5)&(3.4,0.4)&(1.8,0.2)&(2.0,0.2)&(2.1,0.2)&(0.9,0.1)&(1.0,0.1)&(70.1,3.2)&(12.3,0.2)&(13.7,0.0)&&&&7.5\\
4619.7&(9.4,1.2)&(4.5,0.5)&(4.7,0.5)&(0.0,0.0)&(5.1,0.5)&(4.8,0.4)&(0.0,0.0)&(6.0,0.3)&(5.4,0.3)&(0.0,$-$)&(47.1,$-$)&(46.2,$-$)&&&&3.7\\
4605.1&(0.0,0.0)&(6.5,0.7)&(6.4,0.6)&(0.1,0.0)&(0.1,0.0)&(0.0,0.0)&(80.8,4.6)&(18.7,0.7)&(20.7,0.8)&(0.1,$-$)&(0.0,$-$)&(0.0,$-$)&&&&7.4\\
4568.1&(7.0,0.8)&(3.6,0.3)&(3.9,0.3)&(0.0,0.0)&(1.4,0.1)&(1.3,0.1)&(0.0,0.0)&(57.7,$-$)&(56.3,$-$)&(0.0,$-$)&(1.1,$-$)&(1.1,$-$)&&&&1.6\\
4503.0&(0.0,0.0)&(14.6,0.7)&(12.7,0.6)&(70.7,2.7)&(15.1,$-$)&(18.3,$-$)&(1.1,$-$)&(0.3,$-$)&(0.4,$-$)&(0.0,$-$)&(0.0,$-$)&(0.0,$-$)&&&&4.0\\
4468.0&(6.6,0.5)&(16.1,$-$)&(18.0,$-$)&(0.1,$-$)&(47.7,$-$)&(44.4,$-$)&(0.0,$-$)&(0.3,$-$)&(0.2,$-$)&(0.0,$-$)&(0.0,$-$)&(0.0,$-$)&&&&0.5\\
4405.9&(65.8,1.7)&(30.0,$-$)&(30.3,$-$)&(0.0,$-$)&(3.2,$-$)&(3.3,$-$)&(0.0,$-$)&(0.0,$-$)&(0.0,$-$)&(0.0,$-$)&(0.3,$-$)&(0.3,$-$)&&&&1.7\\
\hline
$J=1/2$&$\Lambda_c^+D_s^-$&$\Xi_c^0\bar{D}^0$&$\Xi_c^+D^-$&$\Lambda_c^+D_s^{*-}$&$\Sigma_c^+D_s^-$&$\Xi_c^{\prime0}\bar{D}^0$&$\Xi_c^{\prime+}D^-$&$\Xi_c^0\bar{D}^{*0}$&$\Xi_c^+D^{*-}$&$\Sigma_c^+D_s^{*-}$&$\Xi_c^{\prime0}\bar{D}^{*0}$&$\Xi_c^{\prime+}D^{*-}$&$\Sigma_c^{*+}D_s^{*-}$&$\Xi_c^{*0}\bar{D}^{*0}$&$\Xi_c^{*+}D^{*-}$&$\Gamma_{sum}$\\
4783.7&(0.0,0.0)&(0.0,0.0)&(0.0,0.0)&(0.0,0.0)&(0.9,0.1)&(0.9,0.1)&(0.9,0.1)&(0.0,0.0)&(0.0,0.0)&(6.9,0.9)&(6.8,0.8)&(6.8,0.8)&(34.7,3.6)&(37.7,3.6)&(37.6,3.6)&13.7\\
4658.6&(0.0,0.0)&(0.0,0.0)&(0.0,0.0)&(0.0,0.0)&(13.6,1.8)&(13.9,1.8)&(14.0,1.8)&(0.1,0.0)&(0.1,0.0)&(24.2,2.1)&(29.6,2.2)&(29.3,2.2)&(1.8,0.1)&(3.4,0.1)&(3.3,0.0)&12.0\\
4622.0&(0.0,0.0)&(1.7,0.3)&(1.7,0.2)&(0.0,0.0)&(0.3,0.0)&(0.4,0.1)&(0.5,0.1)&(23.3,2.5)&(21.8,2.3)&(29.0,1.9)&(4.7,0.3)&(5.4,0.3)&(32.0,$-$)&(5.9,$-$)&(6.6,$-$)&7.9\\
4577.3&(1.5,0.2)&(0.6,0.1)&(0.7,0.1)&(19.5,2.3)&(0.0,0.0)&(1.5,0.2)&(1.4,0.1)&(11.3,1.0)&(12.2,1.1)&(0.0,0.0)&(20.9,$-$)&(20.5,$-$)&(0.0,$-$)&(21.9,$-$)&(21.3,$-$)&5.2\\
4521.0&(0.0,0.0)&(32.0,4.0)&(31.1,3.8)&(0.0,0.0)&(0.4,0.0)&(0.1,0.0)&(0.1,0.0)&(0.9,0.1)&(0.9,0.1)&(27.7,$-$)&(6.3,$-$)&(6.6,$-$)&(19.0,$-$)&(4.0,$-$)&(4.3,$-$)&7.9\\
4465.4&(34.3,4.6)&(14.5,1.5)&(15.4,1.6)&(0.2,0.0)&(0.0,0.0)&(0.4,0.0)&(0.4,0.0)&(0.7,$-$)&(0.7,$-$)&(0.0,$-$)&(19.1,$-$)&(18.7,$-$)&(0.0,$-$)&(14.6,$-$)&(14.3,$-$)&7.8\\
4447.9&(0.0,0.0)&(0.1,0.0)&(0.1,0.0)&(0.0,0.0)&(73.6,3.6)&(15.8,0.3)&(20.1,0.2)&(11.2,$-$)&(8.4,$-$)&(1.2,$-$)&(0.3,$-$)&(0.4,$-$)&(1.4,$-$)&(0.3,$-$)&(0.5,$-$)&4.1\\
4425.3&(0.1,0.0)&(0.0,0.0)&(0.0,0.0)&(7.4,0.4)&(0.1,0.0)&(27.4,$-$)&(23.9,$-$)&(33.9,$-$)&(37.2,$-$)&(0.0,$-$)&(1.0,$-$)&(1.0,$-$)&(0.0,$-$)&(0.7,$-$)&(0.6,$-$)&0.4\\
4397.2&(0.0,0.0)&(0.3,0.0)&(0.3,0.0)&(61.5,$-$)&(0.0,$-$)&(28.4,$-$)&(27.7,$-$)&(7.5,$-$)&(7.3,$-$)&(0.0,$-$)&(0.0,$-$)&(0.0,$-$)&(0.0,$-$)&(0.1,$-$)&(0.1,$-$)&0.0\\
4214.6&(53.0,$-$)&(39.6,$-$)&(39.7,$-$)&(0.2,$-$)&(0.0,$-$)&(0.0,$-$)&(0.0,$-$)&(0.1,$-$)&(0.1,$-$)&(0.0,$-$)&(0.1,$-$)&(0.1,$-$)&(0.0,$-$)&(0.2,$-$)&(0.2,$-$)&0.0\\
\hline
\end{tabular}
\end{sidewaystable}

\begin{sidewaystable}\centering
\caption{Rearrangement decays for the $S$-wave $(qqc)(q\bar{c})$ ($q=u,d,s$) molecular states with strangeness=-1 in the isospin-symmetric case. The numbers in the parentheses are ($100 |{\cal M}|^2/{\alpha^\prime}^2$, $10^7\Gamma/{\alpha^\prime}^2\cdot$MeV). The symbol ``$-$'' means that the decay is forbidden. We do not show open-charm decay channels since their couplings with the decay channels vanish.}\label{decay-uds-mole}
 \resizebox{\textwidth}{60mm}{
\begin{tabular}{c|ccccccccccccccccccccc}\hline
States &\multicolumn{20}{c}{Channels}&$\Gamma_{sum}$\\\hline
$J^P=\frac52^-$&$\Sigma^{*0}J/\psi$&$(\Sigma_{c}^{*+}D_{s}^{*-}$&$\Xi_{c}^{*0}\bar{D}^{*0}$&$\Xi_{c}^{*+}D^{*-})$&&&&&&&&&&&&&&&&&\\
$(\Xi_{c}^{*}\bar{D}^{*})^{I=1}$&(11.1,1.2)&(11.1,0.5)&(50.0,$-$)&(50.0,$-$)&&&&&&&&&&&&&&&&&1.7\\
$(\Xi_{c}^{*}\bar{D}^{*})^{I=0}$&(0.0,$-$)&(0.0,$-$)&(50.0,$-$)&(50.0,$-$)&&&&&&&&&&&&&&&&&0.0\\
$(\Sigma_{c}^{*}\bar{D}_{s}^{*})^{I=1}$&(11.1,1.1)&(100.0,$-$)&(5.6,$-$)&(5.6,$-$)&&&&&&&&&&&&&&&&&1.1\\
\hline
$J^P=\frac32^-$&$\Lambda^{0}J/\psi$&$\Sigma^{0}J/\psi$&$\Sigma^{*0}\eta_{c}$&$\Sigma^{*0}J/\psi$&$(\Lambda_{c}^{+}D_{s}^{*-}$&$\Xi_{c}^{0}\bar{D}^{*0}$&$\Xi_{c}^{+}D^{*-})$&$(\Sigma_{c}^{*+}D_{s}^{-}$&$\Xi_{c}^{*0}\bar{D}^{0}$&$\Xi_{c}^{*+}D^{-})$&$(\Sigma_{c}^{+}D_{s}^{*-}$&$\Xi_{c}^{'0}\bar{D}^{*0}$&$\Xi_{c}^{'+}D^{*-})$&$(\Sigma_{c}^{*+}D_{s}^{*-}$&$\Xi_{c}^{*0}\bar{D}^{*0}$&$\Xi_{c}^{*+}D^{*-})$\\
$(\Xi_{c}^{*}\bar{D}^{*})^{I=1}$&(0.0,$-$)&(1.5,0.2)&(4.6,0.6)&(0.3,0.0)&(0.0,$-$)&(0.0,$-$)&(0.0,$-$)&(4.6,0.5)&(0.0,$-$)&(0.0,$-$)&(1.5,0.1)&(0.0,$-$)&(0.0,$-$)&(0.3,0.0)&(50.0,$-$)&(50.0,$-$)&&&&&1.4\\
$(\Xi_{c}^{*}\bar{D}^{*})^{I=0}$&(4.6,0.8)&(0.0,$-$)&(0.0,$-$)&(0.0,$-$)&(4.6,0.7)&(0.0,$-$)&(0.0,$-$)&(0.0,$-$)&(0.0,$-$)&(0.0,$-$)&(0.0,$-$)&(0.0,$-$)&(0.0,$-$)&(0.0,$-$)&(50.0,$-$)&(50.0,$-$)&&&&&1.4\\
$(\Sigma_{c}^{*}\bar{D}_{s}^{*})^{I=1}$&(0.0,$-$)&(6.2,0.9)&(4.6,0.6)&(0.3,0.0)&(0.0,$-$)&(2.3,0.3)&(2.3,0.3)&(0.0,$-$)&(2.3,0.2)&(2.3,0.2)&(0.0,$-$)&(0.8,0.1)&(0.8,0.0)&(100.0,$-$)&(0.2,$-$)&(0.2,$-$)&&&&&2.5\\
$(\Xi_{c}^{'}\bar{D}^{*})^{I=1}$&(0.0,$-$)&(0.3,0.0)&(3.7,0.5)&(6.2,0.5)&(0.0,$-$)&(0.0,$-$)&(0.0,$-$)&(0.9,0.1)&(0.0,$-$)&(0.0,$-$)&(7.7,0.3)&(50.0,$-$)&(50.0,$-$)&(1.5,$-$)&(0.0,$-$)&(0.0,$-$)&&&&&1.3\\
$(\Xi_{c}^{'}\bar{D}^{*})^{I=0}$&(0.9,0.1)&(0.0,$-$)&(0.0,$-$)&(0.0,$-$)&(0.9,0.1)&(0.0,$-$)&(0.0,$-$)&(0.0,$-$)&(0.0,$-$)&(0.0,$-$)&(0.0,$-$)&(50.0,$-$)&(50.0,$-$)&(0.0,$-$)&(0.0,$-$)&(0.0,$-$)&&&&&0.2\\
$(\Sigma_{c}\bar{D}_{s}^{*})^{I=1}$&(0.0,$-$)&(1.2,0.2)&(3.7,0.4)&(6.2,0.5)&(0.0,$-$)&(0.5,0.0)&(0.5,0.0)&(0.0,$-$)&(0.5,0.0)&(0.5,0.0)&(100.0,$-$)&(3.9,$-$)&(3.9,$-$)&(0.0,$-$)&(0.8,$-$)&(0.8,$-$)&&&&&1.1\\
$(\Xi_{c}^{*}\bar{D})^{I=1}$&(0.0,$-$)&(0.9,0.1)&(2.8,0.3)&(4.6,0.2)&(0.0,$-$)&(0.0,$-$)&(0.0,$-$)&(2.8,0.1)&(50.0,$-$)&(50.0,$-$)&(0.9,$-$)&(0.0,$-$)&(0.0,$-$)&(4.6,$-$)&(0.0,$-$)&(0.0,$-$)&&&&&0.7\\
$(\Xi_{c}^{*}\bar{D})^{I=0}$&(2.8,0.4)&(0.0,$-$)&(0.0,$-$)&(0.0,$-$)&(2.8,0.3)&(0.0,$-$)&(0.0,$-$)&(0.0,$-$)&(50.0,$-$)&(50.0,$-$)&(0.0,$-$)&(0.0,$-$)&(0.0,$-$)&(0.0,$-$)&(0.0,$-$)&(0.0,$-$)&&&&&0.7\\
$(\Sigma_{c}^{*}\bar{D}_{s})^{I=1}$&(0.0,$-$)&(3.7,0.4)&(2.8,0.3)&(4.6,0.1)&(0.0,$-$)&(1.4,0.0)&(1.4,0.0)&(100.0,$-$)&(1.4,$-$)&(1.4,$-$)&(0.0,$-$)&(0.5,$-$)&(0.5,$-$)&(0.0,$-$)&(2.3,$-$)&(2.3,$-$)&&&&&0.7\\
$(\Xi_{c}\bar{D}^{*})^{I=1}$&(0.0,$-$)&(8.3,1.0)&(0.0,$-$)&(0.0,$-$)&(0.0,$-$)&(50.0,$-$)&(50.0,$-$)&(2.8,$-$)&(0.0,$-$)&(0.0,$-$)&(0.9,$-$)&(0.0,$-$)&(0.0,$-$)&(4.6,$-$)&(0.0,$-$)&(0.0,$-$)&&&&&1.0\\
$(\Xi_{c}\bar{D}^{*})^{I=0}$&(2.8,0.4)&(0.0,$-$)&(0.0,$-$)&(0.0,$-$)&(2.8,0.2)&(50.0,$-$)&(50.0,$-$)&(0.0,$-$)&(0.0,$-$)&(0.0,$-$)&(0.0,$-$)&(0.0,$-$)&(0.0,$-$)&(0.0,$-$)&(0.0,$-$)&(0.0,$-$)&&&&&0.6\\
$(\Lambda_{c}\bar{D}_{s}^{*})^{I=0}$&(11.1,1.3)&(0.0,$-$)&(0.0,$-$)&(0.0,$-$)&(100.0,$-$)&(1.4,$-$)&(1.4,$-$)&(0.0,$-$)&(1.4,$-$)&(1.4,$-$)&(0.0,$-$)&(0.5,$-$)&(0.5,$-$)&(0.0,$-$)&(2.3,$-$)&(2.3,$-$)&&&&&1.3\\
\hline
$J^P=\frac12^-$&$\Lambda^{0}\eta_{c}$&$\Sigma^{0}\eta_{c}$&$\Lambda^{0}J/\psi$&$\Sigma^{0}J/\psi$&$\Sigma^{*0}J/\psi$&$(\Lambda_{c}^{+}D_{s}^{-}$&$\Xi_{c}^{0}\bar{D}^{0}$&$\Xi_{c}^{+}D^{-})$&$(\Lambda_{c}^{+}D_{s}^{*-}$&$\Xi_{c}^{0}\bar{D}^{*0}$&$\Xi_{c}^{+}D^{*-})$&$(\Sigma_{c}^{+}D_{s}^{-}$&$\Xi_{c}^{'0}\bar{D}^{0}$&$\Xi_{c}^{'+}D^{-})$&$(\Sigma_{c}^{+}D_{s}^{*-}$&$\Xi_{c}^{'0}\bar{D}^{*0}$&$\Xi_{c}^{'+}D^{*-})$&$(\Sigma_{c}^{*+}D_{s}^{*-}$&$\Xi_{c}^{*0}\bar{D}^{*0}$&$\Xi_{c}^{*+}D^{*-})$\\
$(\Xi_{c}^{*}\bar{D}^{*})^{I=1}$&(0.0,$-$)&(1.9,0.3)&(0.0,$-$)&(0.6,0.1)&(1.2,0.1)&(0.0,$-$)&(0.0,$-$)&(0.0,$-$)&(0.0,$-$)&(0.0,$-$)&(0.0,$-$)&(1.9,0.3)&(0.0,$-$)&(0.0,$-$)&(0.6,0.1)&(0.0,$-$)&(0.0,$-$)&(1.2,0.1)&(50.0,$-$)&(50.0,$-$)&0.6\\
$(\Xi_{c}^{*}\bar{D}^{*})^{I=0}$&(5.6,1.0)&(0.0,$-$)&(1.9,0.3)&(0.0,$-$)&(0.0,$-$)&(5.6,1.0)&(0.0,$-$)&(0.0,$-$)&(1.9,0.3)&(0.0,$-$)&(0.0,$-$)&(0.0,$-$)&(0.0,$-$)&(0.0,$-$)&(0.0,$-$)&(0.0,$-$)&(0.0,$-$)&(0.0,$-$)&(50.0,$-$)&(50.0,$-$)&2.6\\
$(\Sigma_{c}^{*}\bar{D}_{s}^{*})^{I=1}$&(0.0,$-$)&(7.4,1.3)&(0.0,$-$)&(2.5,0.4)&(1.2,0.1)&(0.0,$-$)&(2.8,0.4)&(2.8,0.4)&(0.0,$-$)&(0.9,0.1)&(0.9,0.1)&(0.0,$-$)&(0.9,0.1)&(0.9,0.1)&(0.0,$-$)&(0.3,0.0)&(0.3,0.0)&(100.0,$-$)&(0.6,$-$)&(0.6,$-$)&3.0\\
$(\Xi_{c}^{'}\bar{D}^{*})^{I=1}$&(0.0,$-$)&(0.2,0.0)&(0.0,$-$)&(1.9,0.3)&(2.5,0.2)&(0.0,$-$)&(0.0,$-$)&(0.0,$-$)&(0.0,$-$)&(0.0,$-$)&(0.0,$-$)&(3.7,0.4)&(0.0,$-$)&(0.0,$-$)&(0.3,0.0)&(50.0,$-$)&(50.0,$-$)&(0.6,$-$)&(0.0,$-$)&(0.0,$-$)&0.9\\
$(\Xi_{c}^{'}\bar{D}^{*})^{I=0}$&(0.7,0.1)&(0.0,$-$)&(5.8,0.9)&(0.0,$-$)&(0.0,$-$)&(2.8,0.5)&(0.0,$-$)&(0.0,$-$)&(3.7,0.5)&(0.0,$-$)&(0.0,$-$)&(0.0,$-$)&(0.0,$-$)&(0.0,$-$)&(0.0,$-$)&(50.0,$-$)&(50.0,$-$)&(0.0,$-$)&(0.0,$-$)&(0.0,$-$)&1.9\\
$(\Sigma_{c}\bar{D}_{s}^{*})^{I=1}$&(0.0,$-$)&(0.9,0.2)&(0.0,$-$)&(7.7,1.1)&(2.5,0.2)&(0.0,$-$)&(1.4,0.2)&(1.4,0.2)&(0.0,$-$)&(1.9,0.2)&(1.9,0.2)&(0.0,$-$)&(1.9,0.2)&(1.9,0.2)&(100.0,$-$)&(0.2,$-$)&(0.2,$-$)&(0.0,$-$)&(0.3,$-$)&(0.3,$-$)&2.5\\
$(\Xi_{c}^{'}\bar{D})^{I=1}$&(0.0,$-$)&(0.7,0.1)&(0.0,$-$)&(0.2,0.0)&(7.4,$-$)&(0.0,$-$)&(0.0,$-$)&(0.0,$-$)&(0.0,$-$)&(0.0,$-$)&(0.0,$-$)&(2.8,0.1)&(50.0,$-$)&(50.0,$-$)&(3.7,$-$)&(0.0,$-$)&(0.0,$-$)&(1.9,$-$)&(0.0,$-$)&(0.0,$-$)&0.1\\
$(\Xi_{c}^{'}\bar{D})^{I=0}$&(2.1,0.3)&(0.0,$-$)&(0.7,0.1)&(0.0,$-$)&(0.0,$-$)&(0.0,$-$)&(0.0,$-$)&(0.0,$-$)&(2.8,0.2)&(0.0,$-$)&(0.0,$-$)&(0.0,$-$)&(50.0,$-$)&(50.0,$-$)&(0.0,$-$)&(0.0,$-$)&(0.0,$-$)&(0.0,$-$)&(0.0,$-$)&(0.0,$-$)&0.5\\
$(\Sigma_{c}\bar{D}_{s})^{I=0}$&(0.0,$-$)&(2.8,0.4)&(0.0,$-$)&(0.9,0.1)&(7.4,$-$)&(0.0,$-$)&(0.0,$-$)&(0.0,$-$)&(0.0,$-$)&(1.4,$-$)&(1.4,$-$)&(100.0,$-$)&(1.4,$-$)&(1.4,$-$)&(0.0,$-$)&(1.9,$-$)&(1.9,$-$)&(0.0,$-$)&(0.9,$-$)&(0.9,$-$)&0.4\\
$(\Xi_{c}\bar{D}^{*})^{I=1}$&(0.0,$-$)&(6.3,0.9)&(0.0,$-$)&(2.1,0.2)&(0.0,$-$)&(0.0,$-$)&(0.0,$-$)&(0.0,$-$)&(0.0,$-$)&(50.0,$-$)&(50.0,$-$)&(2.8,0.2)&(0.0,$-$)&(0.0,$-$)&(3.7,$-$)&(0.0,$-$)&(0.0,$-$)&(1.9,$-$)&(0.0,$-$)&(0.0,$-$)&1.3\\
$(\Xi_{c}\bar{D}^{*})^{I=0}$&(2.1,0.3)&(0.0,$-$)&(0.7,0.1)&(0.0,$-$)&(0.0,$-$)&(0.0,$-$)&(0.0,$-$)&(0.0,$-$)&(2.8,0.2)&(50.0,$-$)&(50.0,$-$)&(0.0,$-$)&(0.0,$-$)&(0.0,$-$)&(0.0,$-$)&(0.0,$-$)&(0.0,$-$)&(0.0,$-$)&(0.0,$-$)&(0.0,$-$)&0.5\\
$(\Lambda_{c}\bar{D}_{s}^{*})^{I=0}$&(8.3,1.2)&(0.0,$-$)&(2.8,0.3)&(0.0,$-$)&(0.0,$-$)&(0.0,$-$)&(0.0,$-$)&(0.0,$-$)&(100.0,$-$)&(1.4,$-$)&(1.4,$-$)&(0.0,$-$)&(1.4,$-$)&(1.4,$-$)&(0.0,$-$)&(1.9,$-$)&(1.9,$-$)&(0.0,$-$)&(0.9,$-$)&(0.9,$-$)&1.5\\
$(\Xi_{c}\bar{D})^{I=1}$&(0.0,$-$)&(2.1,0.2)&(0.0,$-$)&(6.3,0.4)&(0.0,$-$)&(0.0,$-$)&(50.0,$-$)&(50.0,$-$)&(0.0,$-$)&(0.0,$-$)&(0.0,$-$)&(0.0,$-$)&(0.0,$-$)&(0.0,$-$)&(2.8,$-$)&(0.0,$-$)&(0.0,$-$)&(5.6,$-$)&(0.0,$-$)&(0.0,$-$)&0.6\\
$(\Xi_{c}\bar{D})^{I=0}$&(0.7,0.1)&(0.0,$-$)&(2.1,0.2)&(0.0,$-$)&(0.0,$-$)&(2.8,0.2)&(50.0,$-$)&(50.0,$-$)&(0.0,$-$)&(0.0,$-$)&(0.0,$-$)&(0.0,$-$)&(0.0,$-$)&(0.0,$-$)&(0.0,$-$)&(0.0,$-$)&(0.0,$-$)&(0.0,$-$)&(0.0,$-$)&(0.0,$-$)&0.4\\
$(\Lambda_{c}\bar{D}_{s})^{I=0}$&(2.8,0.3)&(0.0,$-$)&(8.3,0.5)&(0.0,$-$)&(0.0,$-$)&(100.0,$-$)&(1.4,$-$)&(1.4,$-$)&(0.0,$-$)&(0.0,$-$)&(0.0,$-$)&(0.0,$-$)&(0.0,$-$)&(0.0,$-$)&(0.0,$-$)&(1.4,$-$)&(1.4,$-$)&(0.0,$-$)&(2.8,$-$)&(2.8,$-$)&0.8\\\hline
\end{tabular}}
\end{sidewaystable}

To test whether the considered pentaquark picture and the adopted models are correct or not, we show our predictions for the open-charm decays of the $(uds)_{8_c}(c\bar{c})_{8_c}$ states in Table \ref{decay-uds-compact}. From the results, the widths of such states should be around tens of MeV, similar to the $(uud)_{8_c}(c\bar{c})_{8_c}$ case, if the same $\alpha$ is adopted. The lowest state around 4.2 GeV is interesting because of its narrow width. Although we cannot give nonvanishing partial widths for its hidden-charm decay mode, in principle, it can decay into $\eta_c\Lambda$ or $\eta_c\Sigma^0$. If our mass is underestimated, it can also decay into $J/\psi\Lambda$. From the coupling strength in Table \ref{decay-uds-compact}, for its open-charm decay mode, the dominant channel should be $\Lambda_c^+D_s^-$ if the kinematics is allowed. This charged mode may also be helpful to search for a broader pentaquark around 4470 MeV ($\approx$ $\Xi_c^0\bar{D}^{*0}$ threshold). Around 4.4 GeV ($\approx$ $\Lambda_c^+D_s^{*-}$ threshold), probably there are two almost degenerate pentaquark states coupling strongly with the $\Lambda_c^+D_s^{*-}$ channel. If one of them is slightly below the threshold and the other is slightly above the threshold, one gets a narrow state and a broader state. Around the $\Sigma_c^+D_s^-$ threshold, a narrow state is possible, although it may be slightly above the threshold. Other predictions can be similarly discussed with the help of Table \ref{decay-uds-compact} and Fig. \ref{fig-uds}. It seems that more exotic structures in the $udsc\bar{c}$ system are possible than those in $uudc\bar{c}$.

For a comparison study, we also estimate the partial widths for $S$-wave molecular $udsc\bar{c}$ states. The results are presented in Table \ref{decay-uds-mole}. Different from the $uudc\bar{c}$ case, not all the open-charm decays vanish in the adopted simple rearrangement model, but the partial widths are comparable to the hidden-charm decay widths, even smaller. It looks like that, in the molecule picture, the hidden-charm channels are more important than the open-charm channels. Whether this is true needs more detailed studies because contributions from channel couplings and spacial wave functions are not considered in the present work. Anyway, the differences in the decay behaviors in the compact picture and the molecule picture are worthwhile study.

\section{Discussions and summary}\label{sec5}

In the $(uud)_{8_c}(c\bar{c})_{8_c}$ pentaquark picture, with the approximation $\Gamma_{tot}\approx \Gamma_{sum}$, one gets only one $J^P$ assignment, $(P_c(4457)^+,P_c(4440)^+,P_c(4312)^+)\to (3/2^-,1/2^-,3/2^-)$, from their masses and the decay widths. In the $S$-wave $(nnc)(n\bar{c})$ ($n=u,d$) molecule picture, with the extracted constraints on the branching ratios into $J/\psi$ from Ref. \cite{Cao:2019kst}, one has the assignment $(P_c(4457)^+,P_c(4440)^+,P_c(4312)^+)\to (3/2^-,1/2^-,1/2^-)$, but the assignment $(P_c(4457)^+,P_c(4440)^+,P_c(4312)^+)\to (1/2^-,3/2^-,1/2^-)$ is also possible. It is easy to identify the angular momenta of the $P_c$ states with their open-charm decay channels. From the results in Table \ref{decay-uds-mole}, the hidden-charm channels may have even larger branching ratios than the open-charm channels in the molecule picture. Therefore, at present, one cannot establish the molecule assignments for these three $P_c$ states without any doubt. More experimental decay information will be helpful to understand their inner structures.

In the pentaquark picture we consider, from Fig. \ref{fig-uud}, two more $J/\psi p$ states around the $P_c(4312)^+$ and two $J/\psi p$ states below the $\Sigma_c^*\bar{D}^*$ threshold are expected. In the molecule picture, one $\Sigma_c^*\bar{D}$ state and three $\Sigma_c^*\bar{D}^*$ states should exist if short-range interactions are not considered. The mass relations between the three $\Sigma_c^*\bar{D}^*$ molecules should be $(\Sigma_c^*\bar{D}^*)_{J=1/2}<(\Sigma_c^*\bar{D}^*)_{J=3/2}<(\Sigma_c^*\bar{D}^*)_{J=5/2}$. In the possibility that the $(\Sigma_c^*\bar{D}^*)_{J=1/2}$ is close to the $\Sigma_c\bar{D}^*$ threshold, one more state around the $P_c(4440)^+$ is expected. If short-range interactions are considered, the number of molecular states will be reduced because of the forbidden states \cite{Yamaguchi:2017zmn}. Therefore, the full $J/\psi p$ spectrum is also helpful to establish state assignments.

Now we take a look at the rough values of coupling constants in the adopted decay model by treating the $P_c(4457)^+$ ($P_c(4440)^+$) as a $J^P=3/2^-$ ($1/2^-$) state. If one uses the measured width of $P_c(4440)^+$, the value of $\alpha$ in the compact pentaquark picture is about 8.4 GeV while that of $\alpha^\prime$ in the molecule picture is around 1 GeV. If one adopts the width of $P_c(4457)^+$, we get $\alpha\sim 7.6$ GeV and $\alpha^\prime\sim 2$ GeV. Because the simple rearrangement model cannot describe the hidden-charm and open-charm decays simultaneously, a smaller $\alpha$ should be more reasonable. It should not be as small as $\alpha^\prime$, otherwise the hidden-charm decay ratios in the pentaquark picture will be contradicted with the constraints extracted in Ref. \cite{Cao:2019kst}. New measurements on branching ratios can help to improve the decay model.

As mentioned in the introduction, the $(qqq)_{1_c}(c\bar{c})_{1_c}$ components are excluded in both the considered pentaquark picture and the molecule picture. If the hidden-color components, the short-range quark exchanges, and the coupled-channel effects are considered in the molecule configuration, these two pictures are equivalent in the color-spin structures. The results in the same quark-level dynamical model should also be similar. However, the hadron-level molecular models may give different results. The finding that the properties of $P_c(4457)^+$ and $P_c(4440)^+$ can be explained in two different pictures is interesting. Since the short-range interactions are emphasised in the pentaquark picture while the long-range meson-exchange interactions are important in the molecule picture, probably the finding implies that the short-range quark-gluon interactions and the long-range boson exchange interactions are both important for these two states. For the $P_c(4312)^+$, it is not clear which picture is the correct one. If short-range interaction and hidden-color contributions are important, it is also possible to understand its properties with $J^P=3/2^-$ in the molecule picture. The measured $J^P$ will be helpful to clarify the underlying interactions.

In the general case, probably the molecule picture including short-range interactions \cite{Yamaguchi:2017zmn} is a more realistic description for the $P_c$ states, where the $(uud)_{8_c}(c\bar{c})_{8_c}$ structure can be viewed as an equivalent description for the short-range core of the $(nnc)(n\bar{c})$ molecules. To some extent, the situation is similar to the $X(3872)$ in the scheme that it is a $c\bar{c}$ state affected by coupled channel effects \cite{Liu:2019zoy}. From the coupling strengths given in Table \ref{decay-uud-compact}, one may understand how short-range coupled channels affect the decay properties in the molecule picture. The $P_c(4457)^+$ state couples mainly with the $\Sigma_c\bar{D}^*$ channels. The $P_c(4440)^+$ couples equally with the $\Sigma_c^*\bar{D}^*$ and $\Sigma_c\bar{D}^*$ modes and couples strongly with the $\Lambda_c\bar{D}^*$ channel. The $P_c(4312)^+$ couples dominantly with the $\Sigma_c^*\bar{D}$ channels as well as the $\Lambda_c\bar{D}^*$. Since the present comparison discussions rely only on simple estimations, deeper understanding is certainly needed.

Based on our calculations, the $J^P$ assignment of $P_c(4312)$ and the branching ratios into $J/\psi p$ of the $P_c$ states provide basic information that can be used to distinguish their inner structures. However, several points should be noted when adopting the above conclusions: i) The central value of the experimental mass difference between $P_c(4440)$ and $P_c(4457)$ is 17 MeV while the number between our $P_c(4440)$ (mass=4451 MeV) and $P_c(4457)$ (mass=4448 MeV) is only 3 MeV (mass order is also reversed). This can be attributed to the uncertainty of the effective coupling parameters $C_{ij}$'s. Their values in multiquark states can be different from those used here which are extracted from the conventional hadrons; ii) In our discussions, we have used the $J/\psi p$ branching ratios from Ref. \cite{Cao:2019kst} because of lack of experimental data. Note that the results of Ref. \cite{Cao:2019kst} are subject to large off-shell effects and the upper limit of ${\cal B}(P_c\to J/\psi p)$ may be changed. If the experimentally measured branching ratios into $J/\psi p$ are significant, e.g. 20\%, the molecule interpretation of the $P_c$ states would be more natural than the $(uud)_{8_c}(c\bar{c})_{8_c}$ structure. In that case, the latter structure is probably not a correct description for the $P_c$ states; iii) If there is only one pentaquark around the $\Sigma_c\bar{D}$ threshold and its $J^P$ are not $3/2^-$, one may also conclude that the discussed compact pentaquark structure would be problematic; and iv) Once the partial decay width of a pentaquark is measured, one may predict the branching ratios of other channels with the simple decay model ($H_{decay}$=const). If the results are far from the measured values, the model needs to be improved or may be refuted.

To summarize, we have updated the estimated masses of the $(uud)_{8_c}(c\bar{c})_{8_c}$ and $(uds)_{8_c}(c\bar{c})_{8_c}$ pentaquark states obtained in Ref. \cite{Wu:2017weo} by considering isospin breaking effects. The rearrangement decay properties are considered within a simple model. Decays in the molecule picture is also crudely discussed. We find that both the masses and decay properties of the $P_c(4457)^+$, $P_c(4440)^+$, and $P_c(4312)^+$ can be understood if one treats them as $J^P=3/2^-$, $1/2^-$, and $3/2^-$ compact pentaquark states, respectively. These properties can also be explained in the molecule picture if one treats them as $J^P=3/2^-$ ($1/2^-$), $1/2^-$ ($3/2^-$), and $1/2^-$ $S$-wave states, respectively. Although both pictures are acceptable, probably the molecular model including short-range contributions and coupled channel effects is more realistic in describing the hidden-charm pentaquark states. Further information from experiments, in particular information about the $P_c(4312)^+$, is needed to distinguish their inner structures. We hope the predictions in this study can help future investigations.

{\it Note added}: After the submission of the paper, more studies about the properties of the LHCb pentaquark states appeared \cite{Wang:2019dsi,Holma:2019lxe,Wu:2019rog,Voloshin:2019aut,Sakai:2019qph,Wang:2019hyc,Xu:2019zme,Yamaguchi:2019seo,Valderrama:2019chc,Liu:2019zvb,Ali:2019clg,Winney:2019edt,Pan:2019skd,Burns:2019iih,Pimikov:2019dyr}. A number of them discussed the $J^P$ assignments of the LHCb pentaquark states. In the molecule picture \cite{Wu:2019rog,Voloshin:2019aut,Sakai:2019qph,Wang:2019hyc,Xu:2019zme,Yamaguchi:2019seo,Valderrama:2019chc,Liu:2019zvb,Pan:2019skd,Burns:2019iih}, the $J^P$ of $P_c(4312)$ are always $1/2^-$ while those of [$P_c(4440), P_c(4457)$] can be $[1/2^-,3/2^-]$, $[3/2^-,1/2^-]$, or $[3/2^-,1/2^+]$. In the diquark-diquark-antidiquark configuration \cite{Ali:2019clg}, the $J^P$ of $P_c(4312)$ can be $3/2^-$ while those of [$P_c(4440), P_c(4457)$] are $[3/2^+,5/2^+]$. A study with QCD sum rule indicates that the lowest $(udc)_{8_c}(\bar{c}u)_{8_c}$ state with mass 4.4 GeV has $J^P=1/2^-$ \cite{Pimikov:2019dyr}.

\section*{Acknowledgements}

This project is supported by National Natural Science Foundation of China under Grants No. 11775132.



\end{document}